%% file: main.tex
\documentclass[sigconf]{acmart}
\usepackage[english]{babel}
\usepackage[utf8x]{inputenc}
\usepackage[T1]{fontenc}
\usepackage{dsfont}

\usepackage{booktabs}
\usepackage{color}
\usepackage{colortbl}
\usepackage{amsmath}
\usepackage{graphicx}
\usepackage{booktabs}
\usepackage{todonotes}
\hypersetup{draft}
\usepackage[bookmarksnumbered,unicode]{hyperref}
\usepackage{subcaption}
\definecolor{Gray}{gray}{0.9}
\usepackage{float}
\restylefloat{table}

\newcommand\fullparbox[2]{\parbox{#1}{\strut #2 \strut}}

\copyrightyear{2019}
\acmYear{2019} 
\setcopyright{iw3c2w3}
\acmConference[WWW '19]{Proceedings of the 2019 World Wide Web Conference}{May 13--17, 2019}{San Francisco, CA, USA}
\acmBooktitle{ Proceedings of the 2019 World Wide Web Conference (WWW'19), May 13--17, 2019, San Francisco, CA, USA}
\acmPrice{}
\acmDOI{10.1145/3308558.3313409}
\acmISBN{978-1-4503-6674-8/19/05}

\title{Rock, Rap, or Reggaeton?: Assessing Mexican Immigrants’ Cultural Assimilation Using Facebook Data}
\author{Ian Stewart}
\thanks{*Study completed while first author was an intern at MPIDR.}
\affiliation{%
  \institution{Georgia Institute of Technology}
}
\email{istewart6@gatech.edu}
\author{Ren\'e D. Flores}
\affiliation{%
  \institution{University of Chicago}
}
\email{renedf@uchicago.edu}
\author{Timothy Riffe}
\affiliation{%
  \institution{Max Planck Institute for Demographic Research}
}
\email{riffe@demogr.mpg.de}
\author{Ingmar Weber}
\affiliation{%
  \institution{Qatar Computing Research Institute}
}
\email{iweber@hbku.edu.qa}
\author{Emilio Zagheni}
\affiliation{%
  \institution{Max Planck Institute for Demographic Research}
}
\email{zagheni@demogr.mpg.de}

\input{abstract}

\input{acm_data}

\begin{document}
\maketitle
\input{intro}
\input{related}
\input{data}
\input{results}
\input{discussion}
\input{acknowledgments}

\bibliographystyle{ACM-Reference-Format}
\bibliography{main}

\end{document}

%% file: abstract.tex
\begin{abstract}
The degree to which Mexican immigrants in the U.S. are assimilating culturally has been widely debated. 
To examine this question, we focus on musical taste, a key symbolic resource that signals the social positions of individuals. 
We adapt an assimilation metric from earlier work to analyze self-reported musical interests among immigrants in Facebook. 
We use the relative levels of interest in  musical genres, where a similarity to the host population in musical preferences is treated as evidence of cultural assimilation. 
Contrary to skeptics of Mexican assimilation, we find significant cultural convergence even among first-generation immigrants, which problematizes their use as assimilative ``benchmarks'' in the literature. 
Further, 2nd generation Mexican Americans show high cultural convergence vis-\`a-vis both Anglos and African-Americans, with the exception of those who speak Spanish. 
Rather than conforming to a single assimilation path, our findings reveal how Mexican immigrants defy simple unilinear theoretical expectations and illuminate their uniquely heterogeneous character.

\end{abstract}

%% file: acm_data.tex
\ccsdesc[300]{Human-centered computing~Empirical studies in collaborative and social computing}
\ccsdesc[300]{Information systems~Online advertising}
\ccsdesc[500]{Applied computing~Sociology}

\keywords{immigration, demography, social networks, online advertising}

%% file: intro.tex
\section{Introduction}


Are immigrants assimilating to U.S.\ culture? 
According to the ``straight-line'' assimilation hypothesis, developed by $20^{th}$-century Chicago sociologists, the more time immigrants spend in the host society, the more culturally similar to natives they become~\cite{park1928,thomas1918}. Early European immigrants seemingly conformed to this expectation. 
By the third generation, most European immigrant groups became largely culturally indistinguishable from Anglo-American natives~\cite{warner1945}. 
Nevertheless, post-1965 immigrants from Latin America and Asia do not appear to be following a single, ``linear'' assimilative path. Instead, their assimilation appears ``segmented''~\cite{zhou1994,waters2005}. 
While advantaged immigrants are still assimilating ``linearly,'' disadvantaged groups are taking a less expected route. 
As they concentrate in poor neighborhoods, their children come in close contact with ``downtrodden'' native minorities like African-Americans and often adopt their cultural norms and styles~\cite{portes2001}.
What path are Mexican immigrants taking?
Critics have questioned their willingness and/or capacity to assimilate to mainstream U.S.\ culture. 
Empirical evidence of their assimilation has been mixed~\cite{telles2008,alba2014}. 
We focus on a key dimension of culture, musical taste, to assess the assimilation trajectories of Mexican Americans. 
Patterns of artistic consumption like musical taste can illuminate important aspects of the assimilation process~\cite{thomas2017}. 
Musical tastes both reflect and reproduce ethnic divides~\cite{trienekens2002}.
Hence, shrinking differences in musical tastes across generations would imply that cultural assimilation is taking place and that ethnic boundaries between groups are blurring.


We analyze the musical preferences of Mexican immigrants, Mexican Americans, non-Hispanic Whites (henceforth ``Anglos''), and African-Americans on Facebook.\footnote{\label{repo_note} All code for data collection and analysis available here: \\ \url{https://github.molgen.mpg.de/istewart/immigrant_assimilation}.}
We find that 1st generation Mexican immigrants have relatively high levels of cultural assimilation.
Further, rather than converging with the musical tastes of \emph{either} Anglos or African-Americans, the 2nd generation Mexican-Americans have substantial cultural similarities with \emph{both} groups, which highlights the manifold character of Mexican assimilation.
Finally, assimilation varies significantly based on demographics including age, language preference, and education level.\footnote{Supplementary material for data collection and analysis available here: \\ \url{https://github.molgen.mpg.de/istewart/immigrant_assimilation/blob/master/writings/immigrant_assimilation_supplement.pdf}.}

%% file: related.tex
\section{Related work}
Sociologists of culture have identified musical tastes as key symbolic resources individuals draw on to negotiate and express their social positions ~\cite{weber1958}. 
According to \citeauthor{bourdieu1984}~\cite{bourdieu1984}, taste ``classifies.'' 
It marks identity and social status. 
While taste for high-status musical genres like classical music or opera may signal a privileged class background, lower-status genres suggest a lower class position~\cite{bryson1996}. 
Cultural differences like musical taste may be the building blocks that reinforce and preserve ethnic boundaries~\cite{barth1969,wimmer2008}. 
According to \citeauthor{telles2008}, ``the extent to which such cultural markers [of ethnicity] persist over generations may signify the extent to which ethnic groups remain viable.'' 

However, few studies have examined the musical taste of Mexican immigrants. 
Using inter-generational survey data, \citeauthor{telles2008} find that Mexican Americans express a taste for U.S.\ Latino music as well as for what they term ``American, black'' genres. 
\citet{thomas2017} also uses survey data and finds that Mexican immigrants in the U.S.\ exhibit a wide array of differences in musical taste vis-à-vis Anglos, but these differences diminish among the U.S.\ born second-generation (particularly after adjusting for class). 
Though insightful, this nascent literature has limitations.
Some of these studies only capture a few locales rather than national trends ~\cite{telles2008}. 
Further, first-generation immigrants are normally used as an ``assimilative benchmark'' to which later generation immigrants are compared~\cite{flores2015}, which ignores the fact that first-generation immigrants may themselves acculturate.
Last, prior studies rely on immigrants’ self reports of cultural taste on surveys, which may not capture their everyday preferences. 

By relying on Facebook data, we are able to address these limitations. 
First, Facebook provides rich, nation-wide data on the expressed cultural preferences of its users, which allow researchers to obtain a broader sample than what is available through surveys.
Second, Facebook's worldwide popularity allows us to compare Mexican immigrants' cultural preferences directly with those in the source-country: Mexicans living in Mexico. 
Third, on Facebook, immigrants spontaneously report the music styles that they prefer, which advertisers can then access to target specific subgroups. 
These data provide a more natural, bottom-up window on immigrant interests than a survey. 
While much of Facebook's data is private, researchers can use the advertising platform to obtain aggregate estimates for customized demographics, which can then be validated with ground-truth data~\cite{araujo2017,rampazzo2018}. 
Using the advertising platform, \citeauthor{dubois2018}~\cite{dubois2018} proposed a method to measure immigrant assimilation via cultural interests on Facebook, 
which we apply to examine musical interests of Mexican immigrants.

%% file: data.tex
\section{Data collection}

\subsection{Audience data}

Facebook's advertising platform allows users to compute an estimated audience size for a proposed advertisement~\cite{kosinski2015}.
The audience can be defined by demographic attributes provided by Facebook, including gender, age, home location, ethnic affinity, and interests. 
Some attributes such as gender are self-reported, while others such as ethnic affinity (e.g.\ ``Hispanic (US - All)'') are inferred. 
Facebook users can report their preferences over multiple domains such as music, food, and sports~\cite{dubois2018}. 
In this paper we focus on musical genres as a marker of assimilation~\cite{thomas2017}.\footnotemark[\getrefnumber{repo_note}]

\subsection{Identification of second-generation Mexican immigrants}
Sociologists typically assess immigrant assimilation by comparing the traits and life outcomes of immigrants across generations. 
The expectation is that the U.S. born children of first-generation immigrants will be significantly more acculturated~\cite{greenman2008}~\cite{thomas2017}. 
In this study, the key demographic dimension of interest is generational status. 
Unfortunately, Facebook's advertising platform does not permit this kind of targeting.
We therefore use different Facebook user targets as proxies for the second or later Mexican-American generation:

\begin{enumerate}
\item Hispanic American non-expats with interest in Mexico, \\ i.e. ``Mexican Americans (Mexico)'';
\item Hispanic Spanish-speaking American non-expats, \\ i.e. ``Mexican Americans (Spanish)''; and
\item Hispanic Americans non-expat located in cities with a high concentration of Mexican-Americans, according to 2010 American Communities Survey, \\ i.e. ``Mexican Americans (communities).''
\end{enumerate}

These methods are validated by comparing the estimated distribution of the population across different demographics (age, education, gender, marital status, and region) against known demographic distributions for second-generation Mexican Americans~\cite{barrera2013}.
The sample mean and standard deviation for each demographic distribution are computed from the 2016 Public Use Microdata Sample (PUMS), following the sampling procedure outlined by \citeauthor{torrieri2014}~\cite{torrieri2014}.
We compute the proportion of the population that falls into a given category $c \in C$ within each demographic group $d$ as follows:

\begin{equation}
demo_{c,d} = \frac{count(c,d)}{\sum_{c' \in C}count(c',d)}
\end{equation}
The demographic proportions are computed over the Mexican Americans in the PUMS data, and the three proxy populations described above (Mexico, Spanish, communities).


The demographic distributions in Figure 1 of the supplementary material show that most of the estimates are close to the survey estimates.
The distributions of marital status, age, and education are slightly different from the ground-truth distributions, due to underlying differences in the Facebook user base as compared to the overall population~\cite{smith2018}. 
This bias cannot be ignored, and we therefore caution readers to view our analysis as exploratory rather than definitive.
Nonetheless, the overall distributions align well: the Kullback-Leibler divergence values~\cite{kullback1951} between the estimated and ground-truth populations are significantly lower than the values expected by chance ($KL=0.240, 0.218, 0.236$ for Spanish-speaking, interest and communities as compared to $KL=0.858, 0.936, 1.29$ by chance based on $N=1000$ random samples from uniform distribution). 
We use all three audience estimates as proxies for 2+ generation Mexican Americans in the rest of the study.

\subsubsection{Population statistics}

We are interested in comparing the rate of assimilation among first and later generations of Mexican immigrants, with respect to the majority racial group (Anglos) and a minority racial group (African Americans).
Our study focuses on five main populations: (i) non-Hispanic white Americans (``Anglos''), (ii) African Americans, (iii) Mexican immigrants to the United States, (iv) Mexican-Americans, and (v) Mexican people living in Mexico.
As mentioned above, we define the Mexican-American population in three different ways: interest in Mexico, speaking Spanish, and living in a predominantly Mexican American community.
Note that the populations are not mutually exclusive: for instance, someone may have both an interest in Mexico and may live in a Mexican-American neighborhood.
These ethnic groups are targeted with Facebook's ``ethnic affinities'' demographics, which has been shown to be reasonably accurate in prior work~\cite{speicher2018}.

\input{population_data_summary_table}

We provide a summary of all populations of interest in the study in \autoref{tab:population_data_summary}.
As expected, the Facebook estimates are generally smaller than the survey estimates in part because of incomplete penetration~\cite{smith2018}.
In addition to independent counts, it is worth noting that the Mexican American populations show relatively low overlap: 36.2\% between what we define `communities' and `interest'; 27.2\% between `interest' and `Spanish'; 32.2\% between `Spanish' and `communities'; and 11.7\% between all three.
This provides further justification for treating the Mexican American populations separately throughout the rest of the study.

\subsection{Collect musical interests}
The study focuses on music as a marker of cultural identity, which requires identifying specific musical interests to use in data collection.
An important advantage of using Facebook as data source is that the interests are defined by users, which provides a bottom-up perspective on cultural interests that is often missed by other data sources such as surveys.
It is important to note that because the interests are user-selected, the cultural preferences expressed may be seen as more of an identity ``performance''~\cite{boyd2007} rather than true preference.

We began with a initial set of Facebook musical genres collected in 2016~\cite{bohn2016}, expanded the list manually for Latin American-specific genres from Wikipedia, expanded the list with suggestions from Facebook's advertising platform (manually filtered for musical genre), then filtered to 741 genres with a worldwide audience size above 100,000 (smaller genres are unreliable for comparing sub-populations).
We provide the complete set of genres $\mathbb{I}$ in the supplementary material.

These interests provide the basis for computing differences and similarities across sub-populations. 
Following the approach of   \citeauthor{dubois2018}~(\citeyear{dubois2018}) our analysis starts by computing \emph{interest ratios}, which represent the proportion of a given population $p$ that is interested in genre $i$:
\begin{equation}
\label{eq:interest_ratio}
I_{p,i} = \frac{count(p,i)}{\sum_{i' \in \mathbb{I}}count(p,i')}
\end{equation}
For example, if 10 users from population $p$ show interest in hip-hop music, 60 in rock music and 30 in rap music, then the interest ratio for hip-hop would be $I_{p,hip-hop}=\frac{10}{100}=0.10$. 
This quantity measures the share of interest in a given genre. 
Note that there is potential for overlap between populations. The interest ratio does not account for that, as we have no way of knowing how many users like both hip-hop and rock music, so we treat them as separate populations.



\subsection{Quantify assimilation}
\label{sec:quantify_assimilation}

The intuition behind the measure of assimilation that we propose is that a population of migrants is considered more assimilated if their interests are similar to the interests of the population at the destination, and dissimilar from the interests of the population at the origin.
If the destination population is highly interested in hip-hop music, the origin population shows little interest in hip-hop, but the expat population from origin to destination is highly interested in hip-hop, we would say that the expat population is highly assimilated to the destination population with respect to hip-hop, meaning that expats match the taste of natives for the specific genre.

To quantify the level of assimilation following~\citeauthor{dubois2018}~\cite{dubois2018}, we compute for each genre the assimilation ratio ($AR$) of ex-pat interest proportion and destination interest proportion:  

\begin{equation}
\label{eq:assimilation_ratio}
AR_{i} = \frac{I_{expat,i}}{I_{dest,i}}
\end{equation}
where $I_{expat,i}$ and $I_{dest,i}$ are the proportions of the ex-pat and destination populations that have indicated an interest in musical genre $i$ (\autoref{eq:interest_ratio}).
A higher $AR$ score indicates that the ex-pat population is more assimilated to the source population's interest in genre $i$.


Before computing $AR$, we identify a subset of musical genres $\hat{\mathbb{I}}$ that are strongly identified with the destination population.
The interests in genres are filtered as follows:

\begin{enumerate}
\item Remove all interests $\mathbb{I}'$ that are more associated with the source population than with the destination population: \\ 
$\mathbb{I}'=\{i \in \mathbb{I}; I_{dest,i} < I_{source,i}\}; \hat{\mathbb{I}} = \mathbb{I} \; \backslash \; \mathbb{I}'$.
\item Compute the difference $\Delta$ between destination and source interest proportions, and remove all interests $\mathbb{I}''$ for which the difference falls below the $p^{th}$ percentile ($p=50$ following~\cite{dubois2018}): \\
$\Delta = I_{dest} - I_{source}; \mathbb{I}''=\{i \in \mathbb{I}; \Delta_{i} \leq P_{p}(\Delta) \}; \hat{\mathbb{I}} = \hat{\mathbb{I}} \; \backslash \; \mathbb{I}''$
\end{enumerate}
This produces a set of interests $\hat{\mathbb{I}}$ that are distinctly more associated with the destination population than with the source population (i.e.\ more ``American'' than ``Mexican''), which we use to evaluate assimilation.
Note that this process is repeated for each separate destination population.

%% file: population_data_summary_table.tex
\begin{table}
\centering
\small
\begin{tabular}{p{2.5cm} r r}
\hline
Population name & Facebook estimate & Population estimate \\ \hline
Anglos & 112,000,000 (59\%) & 189,591,066 \\ \hline
\fullparbox{3cm}{African Americans} & 29,000,000 (78\%) & 37,014,326 \\ \hline
U.S. Hispanics & 14,000,000 (57\%) & 24,695,129 \\ \hline
\fullparbox{4cm}{Mexican-Americans \\ (communities)} & 4,100,000 (18\%) & 23,407,709 \\ \hline
\fullparbox{4cm}{Mexican-Americans \\ (Spanish)} & 1,700,000 (9\%) & 19,380,816 \\ \hline
\fullparbox{4cm}{Mexican-Americans \\ (Mexico)} & 2,900,000 & n/a \\ \hline
Mexican immigrants & 9,400,000 (82\%) & 11,508,371 \\ \hline
Mexicans in Mexico & 62,000,000 (56\%) & 111,375,417 \\ \hline
\end{tabular}
\caption{Summary of all populations relevant to the study.}
\label{tab:population_data_summary}
\end{table}

%% file: results.tex
\section{Results}

\subsection{Verifying cultural assimilation}

In this study, we assume that the musical interests expressed on Facebook can measure cultural assimilation.
To validate that assumption, we compute the assimilation ratio for several immigrant groups and validate against results from the literature~\cite{vigdor2011}.
We query Facebook for the audience counts for the top 100 U.S.\ music interests for all immigrants and natives from the following countries: Canada, China, El Salvador, Guatemala, India, Mexico, Philippines, South Korea, and Vietnam (following~\citet{vigdor2011}). 
We then compute all assimilation scores for each immigrant population
using the top 50\% of interests in the target country as compared with the source country.

\begin{figure}
\centering
\includegraphics[width=0.8\columnwidth]{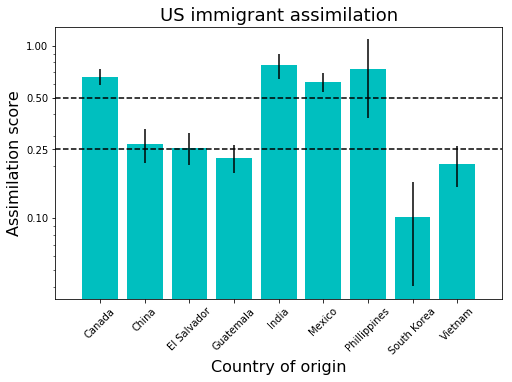}
\caption{Median assimilation scores for selected immigrant groups. Error bars indicate 95\% confidence intervals for the median.}
\label{fig:US_immigrant_assimilation_comparison}
\end{figure}

The median assimilation scores and standard errors are shown in \autoref{fig:US_immigrant_assimilation_comparison}.
We see a clear divide between the high-assimilation immigrant groups (Canada, India, Mexico, and the Philippines) and the low-assimilation immigrant groups (China, El Salvador, Guatemala, South Korea, and Vietnam).
This general trend mirrors the findings of~\citeauthor{vigdor2011}~\cite{vigdor2011} (cf. Figure~6) who found that assimilation for Canadian and Filipino immigrants were higher than for immigrants from China, El Salvador, Guatemala, South Korea, and Vietnam.

While not exhaustive, this test provides some evidence that the behavior of the assimilation scores is consistent with prior work.

\subsection{Comparing assimilation across ethnicity and generations}
Are Mexican immigrants and their descendants becoming culturally assimilated in the U.S.? 
Prior work has typically measured assimilation with respect to the majority U.S. group (``Anglos''). 
However, some scholars have argued that Mexican Americans may be 
acquiring the tastes of marginalized native minorities, such as African Americans~\cite{brubaker2001,thomas2017}.
Therefore, we compare the rate of assimilation across generations with respect to two reference populations: U.S.-born Anglos and African Americans.

\begin{figure}
\centering
\includegraphics[width=\columnwidth]{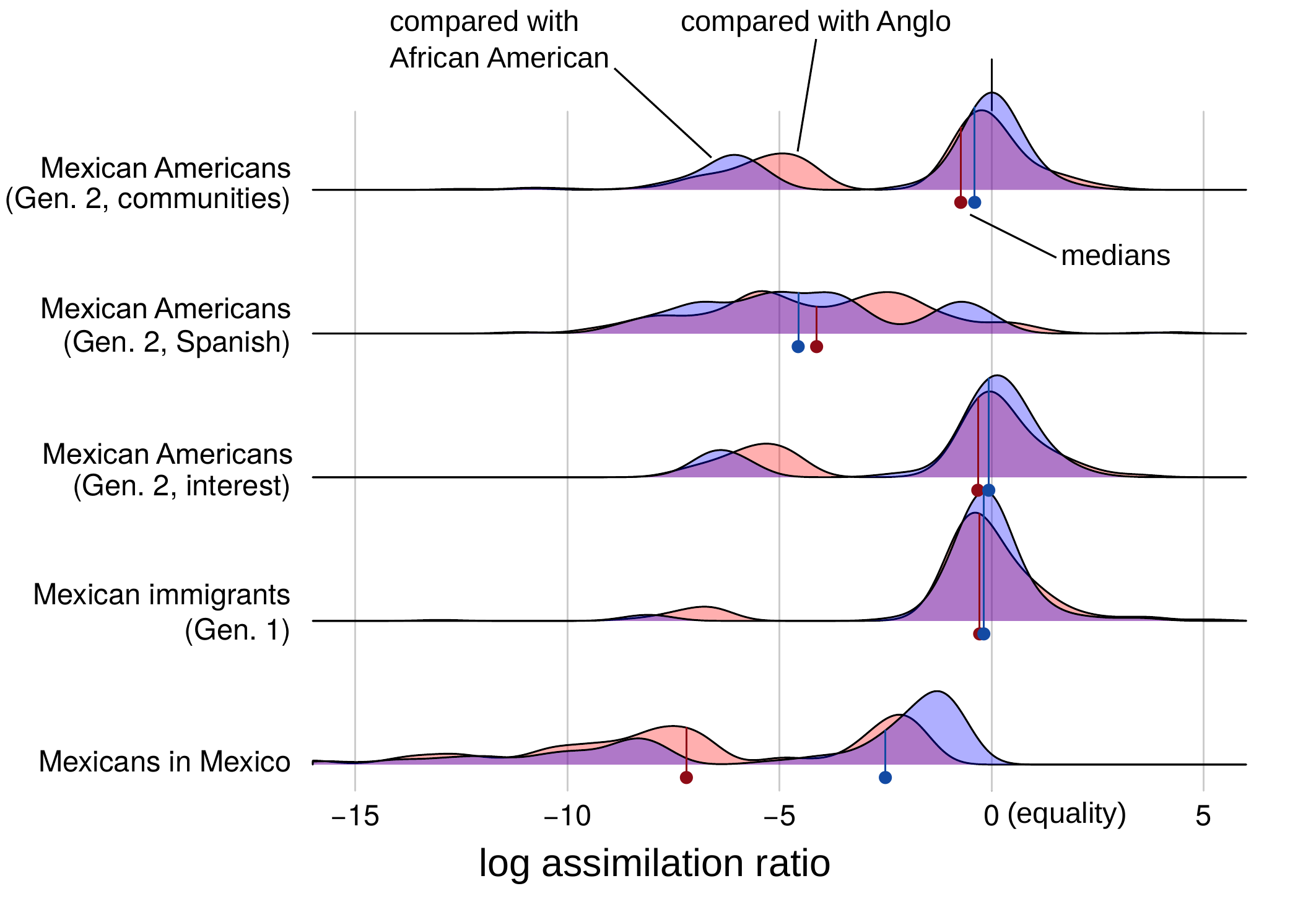}
\caption{Assimilation scores for first- and second-generation Mexican immigrants, separated by target ethnic group (Anglo=red, African-American=blue) and log-transformed.  
Dotted line marks the expected assimilation value $AR=0$, which implies an equal amount of interest for the expat and ethnic group.}
\label{fig:ethnic_assimilation_scores}
\end{figure}

\autoref{fig:ethnic_assimilation_scores} shows the smoothed kernel estimate of assimilation scores between the different Mexican-origin groups and Anglos (in red) and between these groups and the musical tastes of African Americans (in blue). 
It shows that Mexicans in Mexico have the least similar musical tastes relative to Anglos. 
Also, their tastes are more similar to those of African Americans, which may be driven by the growing popularity of rap and hip-hop in Mexico. 
Nevertheless, such greater cultural proximity to African Americans$'$ musical tastes cannot be found among 1st generation Mexican immigrants, who show great convergence with \emph{both} Anglos and Blacks. 
This suggests that Mexican immigrants are a select group and/or that 1st generation immigrants have already acquired U.S. musical tastes during their time in the U.S.

The 2nd generation results are mixed. 
When we measure the Mexican-American 2nd generation by their interests (``interest'') or by their residence (``communities''), we find a high degree of convergence with the musical preferences of \emph{both} Anglos and Blacks, which calls into question unilinear expectations of cultural adaptation. 
In contrast, when we define the 2nd generation based on the ability to speak Spanish, we find greater cultural differences vis-\`a-vis both Anglos and Blacks perhaps because Spanish knowledge is a marker of ``thick'' ethnicity. 
Further, individuals in this group who are not of Mexican origin may be different from their Mexican counterparts.

These results call into question the use of 1st generation immigrants as an “assimilation benchmark” since these immigrants already show substantial cultural assimilation. They also show the sensitivity of assimilation measures to using different definitions of the 2nd generation in Facebook.

\subsection{Comparing assimilation rates across demographics}



Our next analysis investigates patterns of assimilation across different immigrant sub-populations.
We compute the assimilation ratios for the following demographic sub-populations of the Hispanic Mexican ex-pats living in the U.S.:
age, education, gender\footnote{Facebook only allows audience estimates for binary gender, so we restrict our analysis to male versus female while acknowledging that gender is a spectrum.}, dominant language, and geographic region.
These variables are known to modulate the distribution of musical taste among immigrants~\cite{thomas2017} and immigrant assimilation in general~\cite{telles2016}.
With respect to \autoref{eq:assimilation_ratio}, we set the destination population (for $I_{dest}$) to U.S.-born Anglo people and the source population (for $I_{source}$) to Mexico-born people.

\begin{figure}
\centering
\includegraphics[width=\columnwidth]{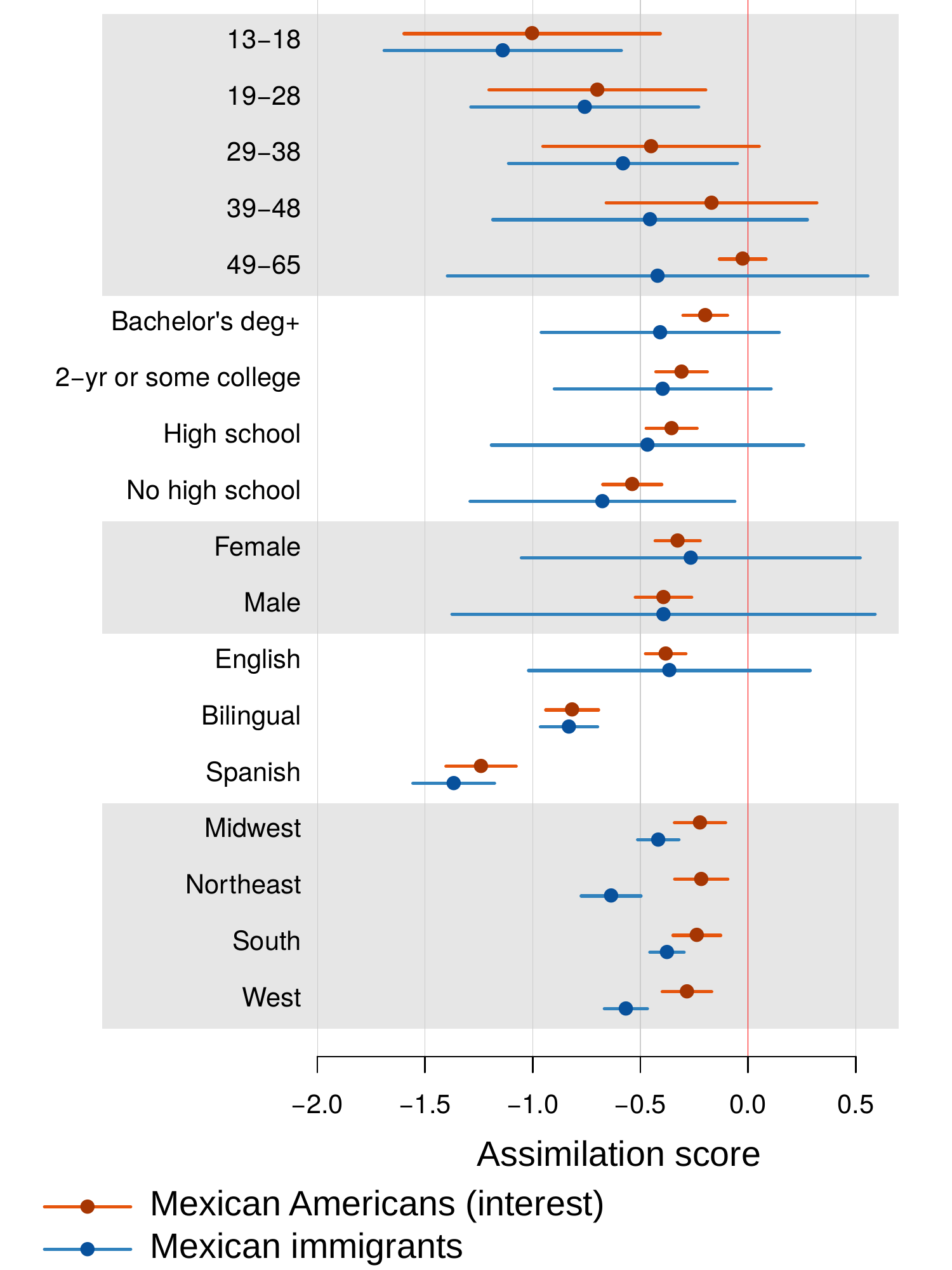}
\caption{Median assimilation scores for Mexican immigrants and Mexican Americans (interest; i.e.\ all Hispanic Americans with a declared interest in Mexico), compared to Anglo interests and subdivided by demographic. The x axis shows log assimilation ratios.}
\label{fig:MX_immigrant_assimilation_scores_by_demographic}
\end{figure}

The distributions in \autoref{fig:MX_immigrant_assimilation_scores_by_demographic} show substantial differentiation among subpopulations (top plot).
There are significant differences for the ages and language demographics (according to Kruskal test, test statistic=$21.6, (p<0.001)$ for age, ($17.6, p<0.001)$ for language):
\begin{enumerate}
\item Younger (13-18) immigrants and Mexican Americans have lower assimilation scores than the older age groups (all groups 29+).
\item English-speaking immigrants and Mexican Americans have higher assimilation scores than Spanish-speaking and Bilingual groups.
\end{enumerate}

The finding about language is more intuitive: English-speaking Facebook users are more likely to understand typical American music (mostly in English).
The age finding is unusual, considering that younger immigrants are more likely to be willing to change their musical interests to align with their peers.
However, analyzing the top-k interests for the youngest age group reveals why they show greater cultural dissimilarity: the top interests for 13-18 Mexican immigrants are highly concentrated under hip-hop and rap genres (``Southern hip hop'', ``Midwest hip hop'', ``Gangsta rap'') and tend to miss the majority of typical Anglo interests (``Alternative rock'', ``Progressive rock'', ``Rock and roll'').

In addition, subgroups of Mexican Americans, especially those defined by educational level, show notable differentiation: 
Mexican Americans with a Bachelor's degree or more tend to have higher assimilation scores than those with less than a high school degree.
Musical taste is not necessarily inaccessible to those with a lower education, so this finding is unanticipated.
In addition, female Mexican Americans have a slight tendency toward higher assimilation than male Mexican Americans, which may be related to different gender socialization among Mexican Americans versus Mexican immigrants.


\subsubsection{Regression analysis}

It is important to note that the prior demographic patterns may be complicated by cross-variable correlations.
For instance, younger Hispanic Facebook users tend to speak more English than older Hispanic Facebook users.
We therefore recompute the assimilation ratios across all cross-sections of the above demographics for all immigrant groups.\footnote{The set of interests $\mathbb{I}$ is restricted to the top 20 music interests to reduce the number of queries.}
We run a linear regression to predict the level of assimilation from the demographic variables:

\begin{align}
AR_{i} = \beta^{T}D_{i} + \epsilon
\end{align}

where $\beta$ is a vector of coefficients for all possible demographic values (male, female, etc.) and $D_{i}$ is a binary vector to indicate the relevant demographic values for group $i$ (male=1, female=0, etc.).
We treat all demographics as categorical even though some may be ordinal (e.g. age, education) to account for possible nonlinear effects.
We run this regression twice, using Anglo and African American Facebook users as the target populations.




The regression results for the Anglo assimilation scores are shown in \autoref{tab:cross_demo_assim_anglo_regression}. 
Some of the earlier results about demographic trends are corroborated here: younger, less educated and English-speaking populations have higher assimilation scores.
We confirm the gender effect outlined earlier, which lends support to the notion of women as leaders of change rather than being culturally conservative.
Lastly, another surprising finding shows a negative correlation with generation: later generations are expected to assimilate more, but after controlling for other variables, all definitions of the later generation have a negative correlation with assimilation. 
This may be due to different distributions of demographics across generations, such as the tendency of the later generation to be younger, which are controlled for in the following regression.

\input{cross_demo_regression_table_anglo}

The regression results for the African American assimilation scores are shown in the supplementary material.
The general patterns remain the same as in the Anglo regression, albeit with some weakened effects (``West'' and ``female'' have lower coefficients).

To test interactions among variables, we re-run the Anglo assimilation score regressions using three separate models:
\begin{enumerate}
\item Model 1: interaction between age and education
\item Model 2: interaction between age and language use
\item Model 3: interaction between age and generation
\end{enumerate}


We omit the table for space (see supplementary material) but summarize the main effects as follows.
In Model 1, higher education correlates with lower assimilation but the effect is less strong for younger (``19-28'') and somewhat older (``39-48'') populations.
In Model 2, older age has no effect on assimilation by itself but older populations that are English-speaking or Bilingual (``39-48 * English'') tend to have higher assimilation scores.
In Model 3, the later generation still tends to have lower assimilation, but younger members of the later generation (``19-28 * Gen. 2'') have higher assimilation.

Across all analyses, we find that several high-level patterns are consistent for Mexican immigrants and Mexican Americans. 
We also find that there are important sources of heterogeneity such as different language effects and interactions between age and other demographics.

\subsection{Summary of findings}

\begin{enumerate}

\item The musical tastes of Mexicans in Mexico are more similar to African Americans’ tastes than to Anglos’.
\item 1st generation Mexican immigrants exhibit high assimilation to Anglos and African Americans.
\item 2nd generation Mexican Americans also exhibit significant assimilation to both Anglos and African Americans with the exception of those that predominantly use Spanish as their main language. 
\item In aggregate, higher-educated, older, and English-speaking populations show more assimilation to Anglo musical tastes. When controlling for cross-demographic effects, female, younger and 1st generation show more assimilation.
\end{enumerate}

%% file: cross_demo_regression_table_anglo.tex
\begin{table}
\centering
\begin{tabular}{l r r}
\toprule
{} & $\beta (S.E.)$ & p \\
\midrule
Intercept & -0.791 (0.054) & *** \\
\textbf{Gender} & ~ & ~ \\
\hspace{8pt}Male & -0.272 (0.027) & *** \\
\textbf{Education} & ~ & ~ \\
\hspace{8pt}High school graduate & -0.001 (0.037) & ~ \\
\hspace{8pt}Less than high school graduate & -0.381 (0.038) & *** \\
\hspace{8pt}Two-year degree, Some college & -0.190 (0.038) & *** \\
\textbf{Age} & ~ & ~ \\
\hspace{8pt}19-28 & 1.168 (0.042) & *** \\
\hspace{8pt}29-38 & 0.996 (0.043) & *** \\
\hspace{8pt}39-48 & 0.878 (0.043) & *** \\
\hspace{8pt}49-65 & 0.730 (0.043) & *** \\
\textbf{Language} & ~ & ~ \\
\hspace{8pt}English & 0.464 (0.033) & *** \\
\hspace{8pt}Spanish & -1.045 (0.035) & *** \\
\textbf{Region} & ~ & ~ \\
\hspace{8pt}Northeast & -0.336 (0.042) & *** \\
\hspace{8pt}South & 0.506 (0.036) & *** \\
\hspace{8pt}West & 0.367 (0.037) & *** \\
\textbf{Gen. 2+ population} & ~ & ~ \\
\hspace{8pt}Mexican American (Spanish) & -1.547 (0.052) & *** \\
\hspace{8pt}Mexican American (communities) & -1.203 (0.036) & *** \\
Mexican American (interest) & -0.922 (0.033) & *** \\
\bottomrule
\end{tabular}
\caption{Regression on log assimilation score with Anglo Americans as the destination population. $N=20,145, R^{2}=0.217, F=399.6 (p<0.001)$. *** $p<0.001$, otherwise $p>0.05$.}
\label{tab:cross_demo_assim_anglo_regression}
\end{table}

%% file: discussion.tex
\section{Discussion}

Critics have questioned the capacity and/or willingness of Mexican immigrants to assimilate culturally to the U.S. 
Nevertheless, we find evidence of substantial cultural assimilation even among $1^{\text{st}}$ generation Mexican immigrants, which calls into question their use as a benchmark population. 
Further, we find that 2+ generation Mexican Americans also show high levels of cultural convergence with regards to Anglos and African Americans, with the exception of Spanish-speaking Mexican Americans. 
This could be evidence that Spanish maintenance is a marker of ``thick'' ethnicity~\cite{hale2004}. 



Rather than conforming to a single assimilative path, our findings highlight how Mexicans immigrants defy simple unilinear theoretical expectations. 
Indeed, their remarkable convergence with both Anglo and Black musical tastes, along with the preservation of some ethnic musical genres, reflect Mexican-Americans' heterogeneous character ~\cite{alba2014}. 
Lastly, the greater similarity in musical preferences of Mexicans in Mexico vis-\`a-vis African Americans complicates interpreting a preference for ``black'' genres as a straightforward indicator of downward assimilation among immigrants.

The primary limitation of the study comes from Facebook's ``black box'' algorithms for inferring attributes such as ethnic affinity. 
Though we cannot validate the classifier outputs on a per-user level, previous work has used data on stocks of migrants~\cite{zagheni2017} and voter registration data~\cite{speicher2018} to show that, by and large, the inference for country of origin and ethnic affinity are quite reliable. 
However, such errors are likely non-random, so we cannot rule out potential for bias. 
Replicating our findings with other music-related data sources, such as `Last.fm' \cite{parketal15icwsm}, could be worth pursuing.

Finally, 
it is important to consider privacy implications of our work. 
Whereas Cambridge Analytica and similar companies (1) use individual level data, and (2) actively try to change user behavior, 
we do neither in our work. 
We only work with anonymous, aggregate data that in terms of content is similar to census-level data.



%% file: acknowledgments.tex
\section{Acknowledgments}

The authors would like to thank the researchers at the Max Planck Institute for Demographic Research for their feedback on an early version of this work in summer 2018; Matheus Araujo and Antoine Dubois for their help with accessing Facebook data; and the anonymous reviewers for their thoughtful feedback.
This work received support from the guest program of the Max Planck Institute for Demographic Research.